\documentclass[12pt,review]{elsarticle}
\usepackage[utf8]{inputenc}
\usepackage[T1]{fontenc}
\usepackage{bm}
\usepackage{amssymb}
\usepackage{amsmath}
\usepackage{empheq} 
\usepackage{caption, subcaption}
\usepackage{multirow}

\DeclareSymbolFontAlphabet{\mathcal}   {symbols}
\DeclareMathAlphabet{\mathcal}{OMS}{cmsy}{m}{n}
\SetMathAlphabet{\mathcal}{bold}{OMS}{cmsy}{b}{n}

\usepackage{tabularx}

\newcolumntype{s}{>{\hsize=.2\hsize}X}
\newcolumntype{V}{>{\hsize=.4\hsize}X}

\newcommand{\vel}{\boldsymbol{\mathfrak{u}}}
\newcommand{\Rey}{\mathcal{R}\mathfrak{e}}
\begin{document}
\title{A Level-Set Immersed Boundary Method for Incompressible Flows at Subcritical Reynolds Numbers}
\author[label1]{Radouan Boukharfane\corref{cor1}}
\cortext[cor1]{Corresponding author}
\ead{radouan.boukharfane@um6p.ma}
\address[label1]{Mohammed VI Polytechnic University (UM6P), MSDA Group, Benguerir, Morocco}
\date{\today}
\begin{abstract}
In this work, a numerical scheme based on a level-set immersed boundary method is employed for the numerical simulation of the flow around two tandem circular cylinders in the subcritical flow regimes.
Three different spacing ratios $\ell/\mathcal{D}$ (where $\ell$ is the center-to-center distance between the two cylinders with $\mathcal{D}$ being the diameter of the cylinders) from $2$ to $4$ is considered.
The instantaneous flow structures, pressure distributions and hydrodynamic forces on two tandem cylinders are analyzed at a Reynolds number of $\Rey=2.2\times 10^4$.
The strategy is based on a combination of a narrow-band accurate conservative level set method and ghost-fluid framework.
In this strategy, the interface is defined as the isocontour of a hyperbolic tangent function, which is advected by the fluid, and then periodically reshaped to enforce the degraded level set function being a signed distance function using a reinitialization equation based on an improved form.
The latter approach takes advantage of a mapping onto a classical distance level set while much better preserving the interface shape.
\end{abstract}
\maketitle
\section{Introduction}
The last decades have seen considerable computational techniques proposed to solve partial differential equations on grids that do not necessarily conform to the shape of fluid--immersed solid interface \cite{boukharfane2018contribution}.
This technique, generically referred to as immersed boundary methods (IBM) \cite{boukharfane2018combined}, has some obvious advantages for approaching technical problems involving moving and deformed interfaces.
Indeed, since it is generally based on fixed and Cartesian grids, it eliminates the difficulty and time-consuming nature of complex automated remeshing/remeshing task and it offers an efficient tool for modeling complex structures.
IBM originates from the pioneering work of \citet{peskin1972flow} to introduce a singular source term to the background flow grid in the vicinity of the solid body, where the interaction between the fluid and the immersed boundary is modeled by a well-chosen discretized approximation of the Dirac--function.
IBM been extensively studied in various applications including benchmark geometries \cite{taira2007immersed}, turbulent flows \cite{iaccarino2003immersed}, fluid-structure interaction (FSI) \cite{huang2007simulation}, multiphase flows \cite{o2018volume}, etc. 

From the numerical standpoint, the method originally introduced by \citet{peskin2002immersed} in studying the blood flow through heart valves and the cardiac mechanics  belongs to the category of front tracking methods, where the interface is represented by connected Lagrangian marker points and the flow requires following Lagrangian markers along the fluid/solid interface.
In a grid-based flow solver, one needs to interpolate kernels for a particular Lagrangian point by solving a
weighted least squares problem where data are sampled to evaluate markers velocity and to spread forces on nearby grid points.
This interpolation process constitutes a possible source of errors, which can lead to poor mass conservation near the body, and therefore resulting  in a fluid leak across the interface.
In this work, a strategy based on front capturing
methods, which makes use of of the level set methods is employed to mimics the effect of immersed solid on fluid flows.
It is noteworthy to mention at this stage that the level set methods have not been originally developed to simulate fluid--structure systems.
The reason is that in its classical application, the level set is the interface between two fluids, rather than between a solid and fluid.
The originality of treating the solid--fluid interface as a level set function is that the Lagrangian interfaces are tracked implicitly and are envisioned as level sets
of a advected function.
Compared to the original IBM, the advantage of this formulation is that it takes advantage of the massive recent development of such method for multiphase flows.
As a matter of fact, it would offer more flexibility in handling multiple objects and their topological changes, or more complex fluid–structure coupling, and it enables more  control of mass conservation.

In the present study, the level set method is used to simulate external bluff-body flow around a circular obstacle.
The considered configuration consists of two circular cylinders in tandem arrangement, which is of
particular interest because of the nature of the interaction of the unsteady wake from the upstream cylinder with the downstream cylinder.
Owing to its relevance to many applications, the tandem cylinder conﬁguration has received considerable attention in both recent experimental studies \cite{ljungkrona1991free,khorrami2007unsteady} as well as numerical simulations \cite{kitagawa2008numerical,frendi2009noise}, and has been designated a benchmark aeroacoustic ﬂow by NASA.
The numerical results have been compared with experimental data and past numerical simulation.
The results indicates the considerable potential of the level-set immersed boundary approach to simulate fluid flow problems in complex configurations of bluff bodies.
The next section presents the simulation setup.
The performance of the computational procedure is then addressed in the fourth section, which is focused on the analysis of the computational model response when applied to the flow around two cylinders in tandem arrangement.

\section{Mathematical formulation}
The general equation of level set function and the incompressible two-phase Navier Stokes equations are presented in this section. 
%
\subsection{Level set equation}
%
The Accurate Conservative Level Set (ACLS) method was proposed by \citet{desjardins2008accurate}.
In this method, the Level Set function varies smoothly from 0 to 1 as an hyperbolic tangent across the fluid-solid interface in a conservative manner as 
\begin{equation}
\psi({\bf x},t)=\frac{1}{2}\left(\tanh\left(\frac{\phi({\bf x},t)}{2\varepsilon}\right)+1\right),
\label{eq:level_set_function}
\end{equation}
where the interface is represented in the form of the 0.5 isocontour of the indicator function $\psi({\bf x},t)$.
Away from the interface the level set scalar is assumed to be a signed distance function to the interface, \textit{i.e.}, $\phi({\bf x}, t)=-d$ inside the immersed solid, $\phi({\bf x}, t)=+d$ inside the fluid, and $\phi({\bf x}, t)=0$ at the fluid-solid interface, where $d$ is the shortest distance from a point $\bm{x}$ to the interface at given time $t$.
$\psi$ takes a value $0$ at regions occupied by the solid and $1$ at the fluid.
In Eq.~\eqref{eq:level_set_function}, $\varepsilon$ is a parameter that dictates the thickness of the proﬁle.
The interface location corresponds to the location of the $\psi({\bf x},0) = 0.5$ isosurface.
The evolution of $\psi({\bf x},t)$ in a free divergence velocity field is given by the advection equation as follows,
\begin{equation}\label{eq:conpsi}
\frac{\partial\psi}{\partial t}+\nabla\cdot(\psi\vel)=0.
\end{equation}
In the classical level set method of \cite{olsson2005conservative}, the geometric parameters associated with the interface, such as interface normal vector, \textit{i.e.}, ${\bf n}$) and interface curvature, \textit{i.e}, ${\kappa}$, are evaluated from the level set function following

\begin{subequations}
\begin{empheq}[left={\empheqlbrace\,}]{align}
& {\bf n} = \frac{\nabla \psi}{\lvert \nabla \psi \rvert}
\label{eq:normal}\\
& \kappa = - \nabla \cdot {\bf n}
\label{eq:curvature}
\end{empheq}
\end{subequations}
Accurate conservative level set (ACLS) method relies on a sharper function $\psi$, which can be seen as a volume fraction with a controlled interface width $2\epsilon$.
It is noteworthy that $\phi$ cannot maintain the property $\lvert \nabla \phi \rvert=1$ during transport.
Therefore, there is no guarantee that the hyperbolic tangent profile $\psi$ will remain unchanged.
This takes the form of local modifications of the interface thickness which can lead to an inaccurate representation of the interface and topology computation.
An relaxation step has to be added to make the diffuse-interface profile at equilibrium by solving in pseudo time \cite{olsson2005conservative}
\begin{equation}\label{eq:conspsi}
\frac{\partial\psi}{\partial\tau}+\nabla\cdot \left(\psi(1-\psi)\mathbf{n}\right)=\nabla\cdot\left(\epsilon\left(\nabla\psi\cdot\mathbf{n}\right)\mathbf{n}\right).
\end{equation}
where $\tau$ is a fictitious time step in which the equation is solved until the initial level set profile is recovered.
Equation \eqref{eq:conspsi} is solved in the pseudo--time $\tau$, highlighting the fact that the reinitialization step is purely a numerical constraint.
This procedure of reinitialization slightly changes the position of the interface on a sub-grid scale.
The reinitialization of the level set profile $\psi$ as described by the equation \eqref{eq:conspsi} in which a compressive flux and a diffusive flux are applied in the direction normal to the interface aims to maintain the interface thickness at a constant value.
The stable and accurate method considered here is the ACLS of \citet{chiodi2017reformulation} with the additional modification of \citet{sahut2020evaluation} where the reinitialization is reformulated to
\begin{equation}
\frac{\partial\psi}{\partial\tau} =  \nabla \cdot \left[ \frac{1}{4 \cosh^2 \left( \frac{\phi_{\mathrm{map}} }{2 \epsilon\left(\boldsymbol{x}\right)}\right)}\left(\nabla \phi_{\mathrm{map}}  \cdot \mathbf{n}_{\mathrm{FMM}}-\mathbf{n}_{\mathrm{FMM}}\cdot \mathbf{n}_{\mathrm{FMM}}  \right) \mathbf{n}_{\mathrm{FMM}}\right]
\label{eq:aclsreinit}
\end{equation}
with $\phi_{\mathrm{map}}\left(\boldsymbol{x},\tau\right) = \epsilon\left(\boldsymbol{x}\right) \log \left( \frac{\psi\left(\boldsymbol{x},\tau\right)}{1-\psi\left(\boldsymbol{x},\tau\right)} \right)$ is the inverse of the conservative level set function.
The terms in equation~(\ref{eq:aclsreinit}) are discretized using second order finite differences while the normal $\mathbf{n}_{\mathrm{FMM}}$ is computed as 
\begin{equation}
\mathbf{n}_{\mathrm{FMM}}\left(\boldsymbol{x},t\right) = \frac{\nabla \phi_{\mathrm{FMM}}\left(\boldsymbol{x},t\right)}{\|\nabla \phi_{\mathrm{FMM}}\left(\boldsymbol{x},t\right)\|}
\label{eq:nfmm}
\end{equation}
with $\phi_{\mathrm{FMM}}$ a distance function computed from a Fast Marching Method algorithm \cite{mccaslin2014localized}.
The construction of $\phi_{\mathrm{FMM}}$ is fundamental in the method as it removes all oscillatory behaviors of $\psi$ in the computation of normals \cite{desjardins2008accurate}.
This leads to the following algorithm for a time step
\begin{enumerate}
\item Advance the interface by solving equation~(\ref{eq:conpsi}) to get $\psi^*$ 
\item Evaluate the signed distance $\phi_{\mathrm{FMM}}$ from the isocontour $\psi^*=0.5$, $\phi_{\mathrm{map}}$ from $\psi^*$ and $\mathbf{n}_{\mathrm{FMM}}$ with Eq.~(\ref{eq:nfmm})
\item Carry out one iteration of Eq.~(\ref{eq:aclsreinit}) to get $\psi^{n+1}$ with $\Delta \tau = 0.25\Delta x$
\end{enumerate}
In the ACLS method, the interface mean curvature $\kappa$ is estimated from $\phi$ using Goldman's formula \cite{goldman2005curvature} as follows
\begin{equation}
\kappa = \frac{\bm{\nabla}\phi^\intercal\cdot\boldsymbol{\mathcal{H}}(\phi)\cdot\bm{\nabla}\phi-\|\bm{\nabla}\phi\|^2\mathrm{Tr}\left(\boldsymbol{\mathcal{H}}(\phi)\right)}{\|\bm{\nabla}\phi\|^3}
\end{equation}
where $\boldsymbol{\mathcal{H}}(\phi)$ is the hessian matrix of the signed-distance function given by 
\begin{equation}
\boldsymbol{\mathcal{H}}(\phi)=\begin{pmatrix} 
\frac{\partial^2\phi}{\partial x_1\partial x_1} & \frac{\partial^2\phi}{\partial x_1\partial x_2} & \frac{\partial^2\phi}{\partial x_1\partial x_3} \\ 
\frac{\partial^2\phi}{\partial x_2\partial x_1} & \frac{\partial^2\phi}{\partial x_2\partial x_2} & \frac{\partial^2\phi}{\partial x_2\partial x_3} \\ 
\frac{\partial^2\phi}{\partial x_3\partial x_1} & \frac{\partial^2\phi}{\partial x_3\partial x_2} & \frac{\partial^2\phi}{\partial x_3\partial x_3} 
\end{pmatrix},
\end{equation}
and $\mathrm{Tr}$ is the trace operator given by
\begin{equation}
\mathrm{Tr}\left(\boldsymbol{\mathcal{H}}(\phi)\right)=\frac{\partial^2\phi}{\partial x_1\partial x_1}+\frac{\partial^2\phi}{\partial x_2\partial x_2}+\frac{\partial^2\phi}{\partial x_3\partial x_3}.
\end{equation}
\subsection{Incompressible Navier-Stokes equations with a constant density}

In this study, an incompressible two-phase flow formulation form of the Navier–Stokes equations written in a Cartesian frame of reference $(x_1,x_2,x_3)\equiv (x,y,z)$ is employed,
\begin{equation}\label{eq:consmom}
\frac{\partial\vel}{\partial t}+(\vel\cdot\nabla)\vel=-\frac{1}{\varrho}\nabla\mathfrak{p}+\frac{1}{\varrho}\nabla\cdot\left(\mu\left[\nabla\vel+\nabla\vel^\intercal\right]\right),
\end{equation}
where $\vel$ is the velocity ﬁeld, $\varrho$ is the density, $\mathfrak{p}$ is the pressure, and $\mu$ is the dynamic viscosity.
To ensure that the velocity field is divergence free, the continuity equation is given.
The continuity equation can be written in terms of the incompressibility constraint as
\begin{equation}
\frac{\partial\varrho}{\partial t}+\vel\cdot\nabla\varrho=0
\end{equation}
In the following, the interface separating the two phases is denoted $\Gamma$.
In each phase, the material properties are constant, which allows to write $\varrho=\varrho_s$ in the solid immersed body, while $\varrho=\varrho_f$ in the fluid. 
Similarly, $\mu=\mu_s$ in the solid and $\mu=\mu_f$ in the fluid.
At the interface, the material properties are subject to a jump that is written $\left[\varrho\right]_\Gamma=\varrho_f-\varrho_s$ and $\left[\mu\right]_\Gamma=\mu_f-\mu_s$.
The velocity ﬁeld is continuous across the interface, $\left[\vel\right]_\Gamma=0$.
The existence of the pressure jump induces a discontinuity in the pressure at the interface $\Gamma$, and one can write
\begin{equation}
\left[\mathfrak{p}\right]_\Gamma=\sigma\kappa+\left[\mu\right]_\Gamma\boldsymbol{n}^\intercal\cdot\nabla\vel\cdot\boldsymbol{n}.
\end{equation}
%
\subsection{Ghost-Fluid Method}
The solid--fluid coupling is achieved due to the Ghost-Fluid Method (GFM) \cite{fedkiw1999non}.
It consists in explicitly introducing the singular pressure jump condition into the discretization equations at the interface in the solving of the Poisson equation for the pressure while material discontinuities are accounted for automatically.
The GFM assumes that the jump condition for the pressure $\left[\nabla\mathfrak{p}\right]_\Gamma$ and its spatial derivatives are given at $\Gamma$.
Note that the GFM is based on the extension by continuity of $\nabla\mathfrak{p}_s$ in the solid and of $\nabla\mathfrak{p}_f$ in the fluid.
In a one-dimensional domain, assuming that a node $i$ of the mesh is located in the fluid, and a neighboring node $i+1$ is in the solid, and introducing the index
$\theta= (x_\Gamma-x_i)/(x_{i+1}-x_i)$ and a modified density $\varrho^\ast=\varrho_s\theta+(1-\theta)\varrho_f$, the pressure jump at $x_{i+1}$ is simply computed as
\begin{equation}
\left[\nabla\mathfrak{p}\right]_{i+1}\approx\frac{\varrho_s}{\varrho^\ast}\left[\nabla\mathfrak{p}\right]_\Gamma+\left(1-\frac{\varrho_s}{\varrho^\ast}\right)\left(\mathfrak{p}_{f,i+1}-\mathfrak{p}_{s,i}\right)
\end{equation}
and similar formula to obtain $\left[\nabla\mathfrak{p}\right]_{i}$.
%
%
\subsection{Numerical schemes}
%

The system of equations \eqref{eq:consmom} is solved by means of the projection method based on fractional time steps developed by \citet{chorin1968numerical} and improved by \citet{kim1985application}.
The velocity field is solved at overall iteration times $(n,n+1,\cdots)$, whereas the pressure and the density are solved at half iteration times $(n+\frac12,n+\frac32,\cdots)$.
First, a velocity predictor $\vel^{\ast}$ is computed from Eq.~\eqref{eq:consmom} from which the
pressure gradient at time $n-\frac12$ is dropped, \textit{i.e.},
\begin{equation}
\frac{\vel^\ast-\vel^n}{\Delta t}=-\left(\vel^n\cdot\nabla\right)\vel^n+\frac{1}{\varrho^{n-\frac12}}\nabla\cdot\left(\mu\left[\nabla\vel^n+\left(\nabla\vel^n\right)^\intercal\right]\right),
\end{equation}
The velocity predictor $\vel^{\ast}$ is not necessarily divergence-free. 
Second, a correction of the velocity predictor is performed using the pressure gradient at time $n+\frac12$ as follows
\begin{equation}\label{eq:divfree}
\frac{\vel^{n+1}-\vel^\ast}{\Delta t}=-\frac{1}{\varrho^{n+\frac12}}\nabla\mathfrak{p}^{n+\frac12},
\end{equation}
which has two unknowns, $\vel^{n+1}$ and $\mathfrak{p}^{n+\frac12}$ ($\varrho^{n+\frac12}$ only depends on the phase).
Since $\vel^{n+1}$ is divergence-free, taking the divergence of Eq.~\eqref{eq:divfree}, one obtains the Poisson equation for the updated pressure $\mathfrak{p}^{n+\frac12}$,
\begin{equation}\label{eq:poissoneq}
\nabla\cdot\left(\frac{1}{\varrho^{n+\frac12}}\nabla\mathfrak{p}^{n+\frac12}\right)=\frac{1}{\Delta t}\nabla\cdot\vel^\ast,
\end{equation}
where $\mathfrak{p}^{n+\frac12}$ is the only unknown.
Numerical discretization of the Poisson equation \eqref{eq:poissoneq} leads naturally to a linear system in which the pressure jump at the interface is imposed using the GFM.
Once the updated pressure $\mathfrak{p}^{n+\frac12}$ is known, Eq.~\eqref{eq:divfree} is used to correct $\vel^\ast$ to compute the updated velocity $\vel^{n+1}$.
In this correction, since the pressure gradient is also computed close to the interface, the pressure jump at
the interface is again imposed in the discretization of $\nabla\mathfrak{p}^{n+\frac12}$ in Eq.~\eqref{eq:divfree}.
In this work, a fourth-order central scheme is used for the spatial integration, and a third-order accurate semi-implicit Crank-Nicolson scheme is employed for time integration. 
To solve the Poison equation in Eq.~\eqref{eq:poissoneq}, 
the Livermore's \texttt{Hypre} library \cite{falgout2002hypre} is used with the PCG (pre-conditioned conjugate gradient) method.

\section{Numerical results}
\begin{figure}[ht!]
\centering
\includegraphics[width=0.99\textwidth]{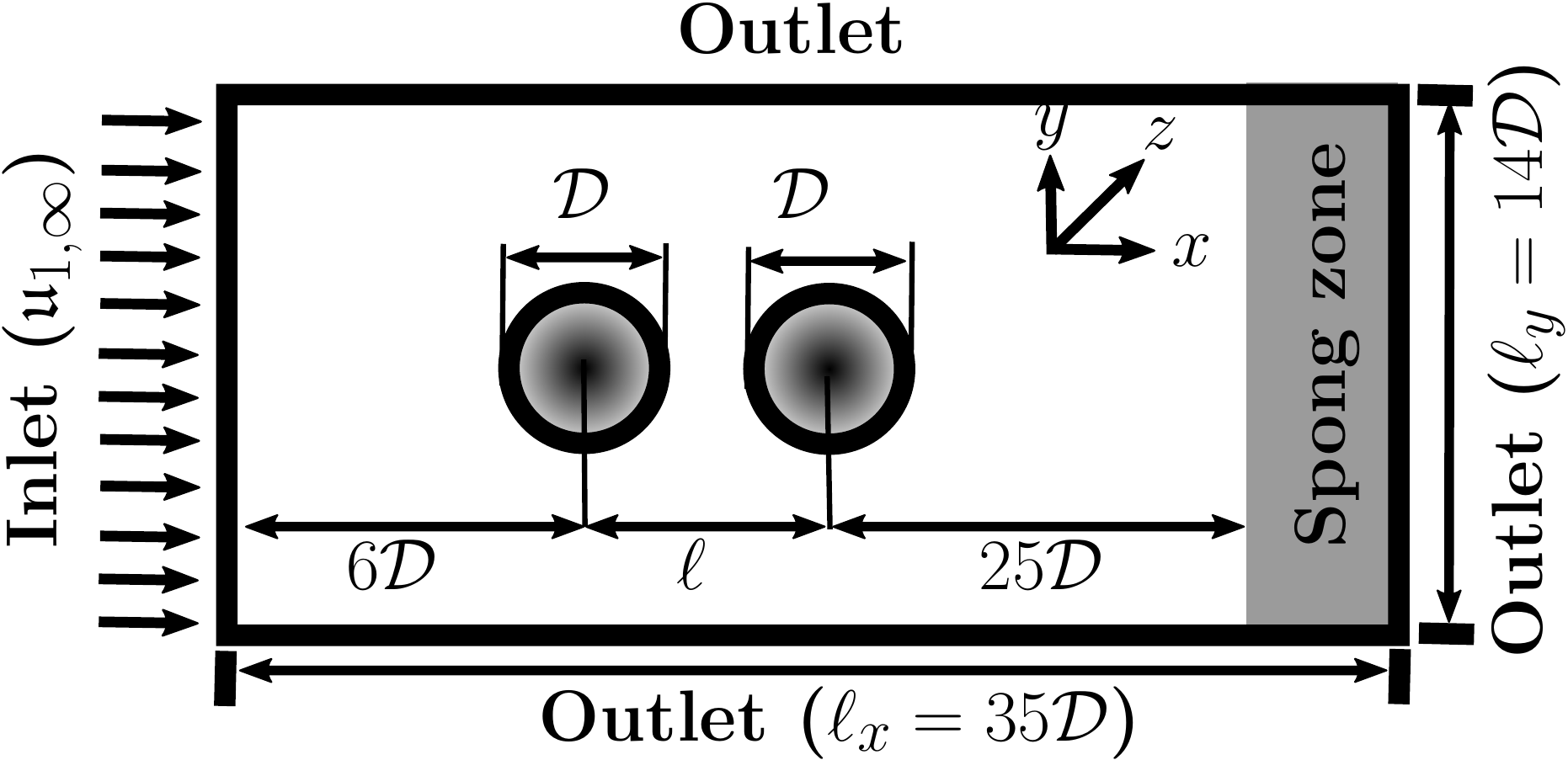}
\caption{Geometric parameters of the simulation domain with two cylinders of different diameters disposed in tandem}
\label{fig:sidesphere}
\end{figure}

The numerical results of an three--dimensional incompressible viscous flows past a pair of circular cylinders in tandem arrangement are performed the asses the validity of the present method.
Hereafter, the Reynolds number is defined as $\Rey=\varrho\mathfrak{u}_{1,\infty}\mathcal{D}/\mu$, in which $\mathfrak{u}_{1,\infty}$ is the free stream velocity and $\mathcal{D}$ is the diameter of the cylinder.
The drag coefficient $\mathcal{C}_d$ and lift coefficient $\mathcal{C}_l$ of cylinder are defined as $\mathcal{C}_d=2\mathcal{F}_d/\varrho\mathfrak{u}_{1,\infty}^2\mathcal{D}$  and $\mathcal{C}_l=2\mathcal{F}_l/\varrho\mathfrak{u}_{1,\infty}^2\mathcal{D}$, where $\mathcal{F}_d$ and $\mathcal{F}_l$ are the drag force and lift force respectively.
The Strouhal number is defined as $\mathrm{St}=\mathfrak{f}_q\mathcal{D}/\mathfrak{u}_{1,\infty}$, where $\mathfrak{f}_q$ is the vortex shedding frequency.
For the all flow past obstacles numerical experiments, the density $\varrho$ of the fluid is set as $1.0$ and to $0.0$ for the solid.
The dynamic viscosity of the fluid is set to $\mu$, while it is set to very higher value for the solid.
In this study, the gap ($\ell$) between the centers of the cylinders is set to $\ell/\mathcal{D}=2$, $3$, and $4$ for each arrangement.
The size of cubic computation domain is $-6\mathcal{D}\le x\le 29\mathcal{D}$, $-7\mathcal{D}\le y\le 7\mathcal{D}$.
For the spanwise domain length, \citet{norberg1994experimental} showed that it is necessary to make $\ell_z/\mathrm{D}\ge 1$ (where $\ell_z$ is the cylinder length) to achieve a good simulation of the
time-mean and fluctuating fluid forces on the cylinder. Thus, following the recommendation of \citet{labbe2007numerical}, the spanwise length $\ell_z$ is set at $\pi\mathrm{D}$ in the present simulation.
The computational domain is discretized using a grid of $1600\times 400\times 120$ points in the $x$, $y$, and $z$ directions, respectively.
In the present computations, a rectangular computational domain is used, as shown in Fig.~\ref{fig:sidesphere}.
The downstream cylinder is located at $6\mathcal{D}$ from the inflow boundary, while the upstream cylinder is positioned at $25\mathcal{D}$ from outgoing boundary.
Note that we use a sponge layer to damp out the possible reflection of pressure fluctuations at the outlet.
The simulations are performed up to the enough
non-dimensional time $t^\ast=\mathfrak{u}_{1,\infty}t/\mathcal{D}$ to acquire stable calculated values.

\section{Results and discussion}
\begin{figure}[ht!]
\centering
\begin{subfigure}[h]{0.99\textwidth}
\centering
\includegraphics[width=0.69\textwidth]{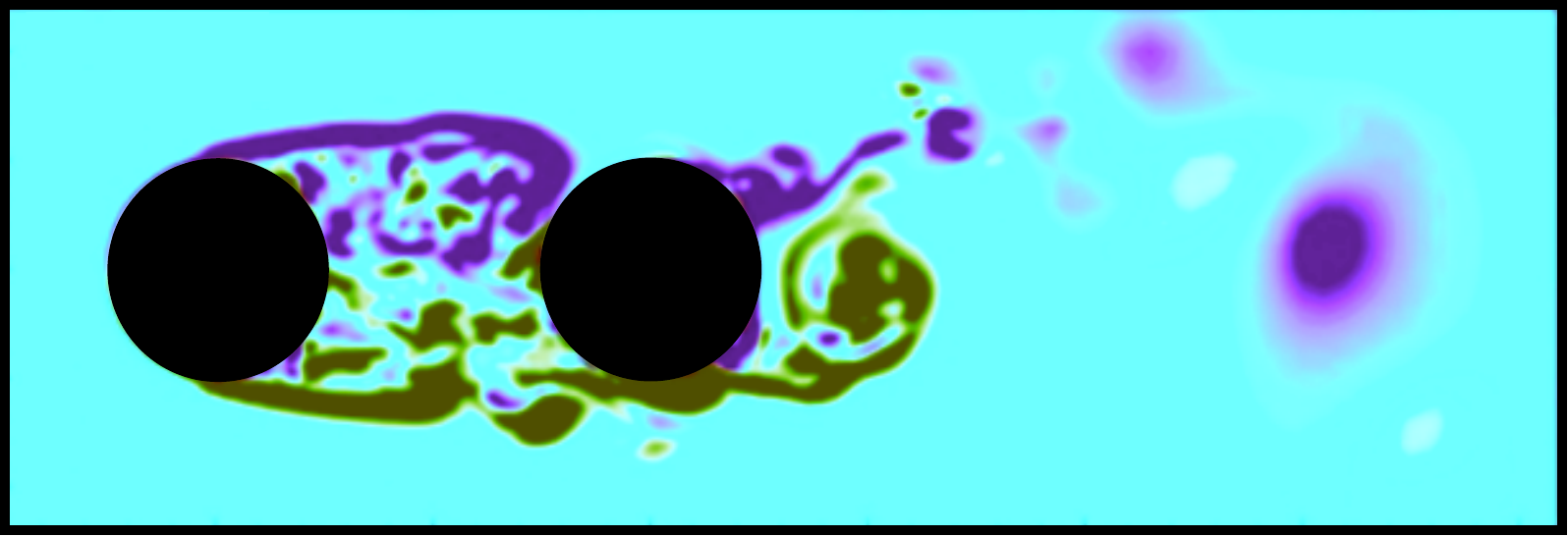}
\caption{$\ell/\mathcal{D}=2$}
\label{fig:midspanvortana}
\end{subfigure}\\
\begin{subfigure}[h]{0.99\textwidth}
\centering
\includegraphics[width=0.69\textwidth]{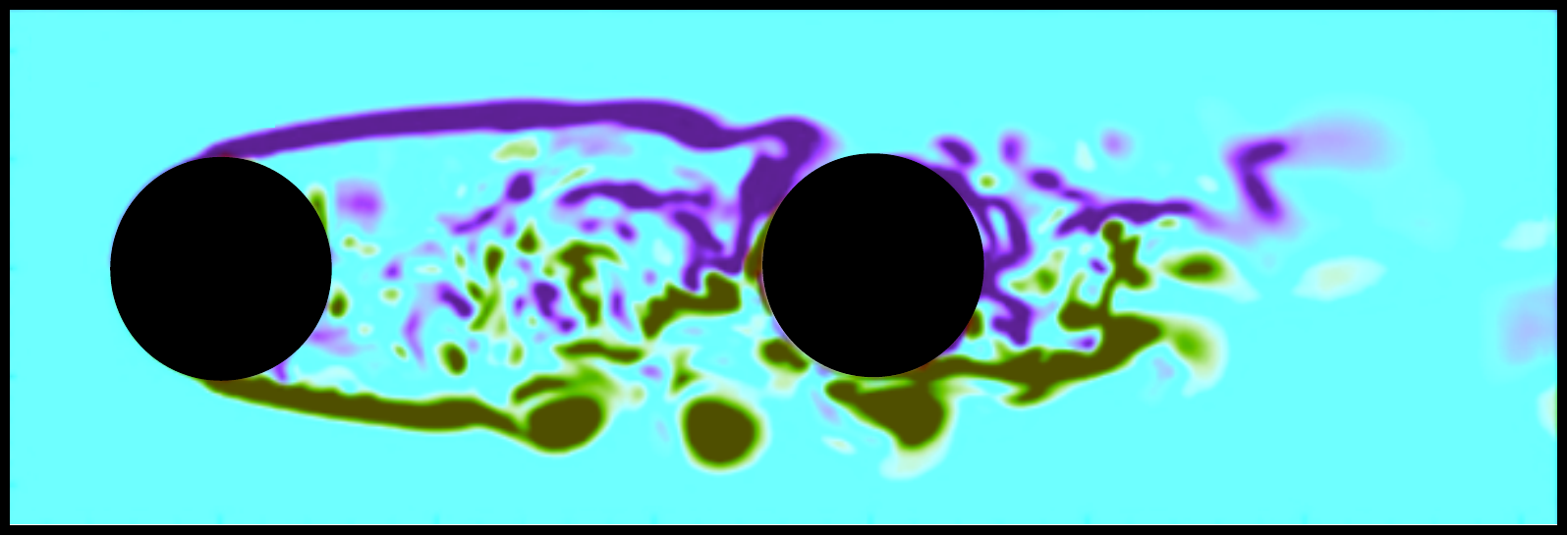}
\caption{$\ell/\mathcal{D}=3$}
\label{fig:midspanvortanb}
\end{subfigure}\\
\begin{subfigure}[h]{0.99\textwidth}
\centering
\includegraphics[width=0.69\textwidth]{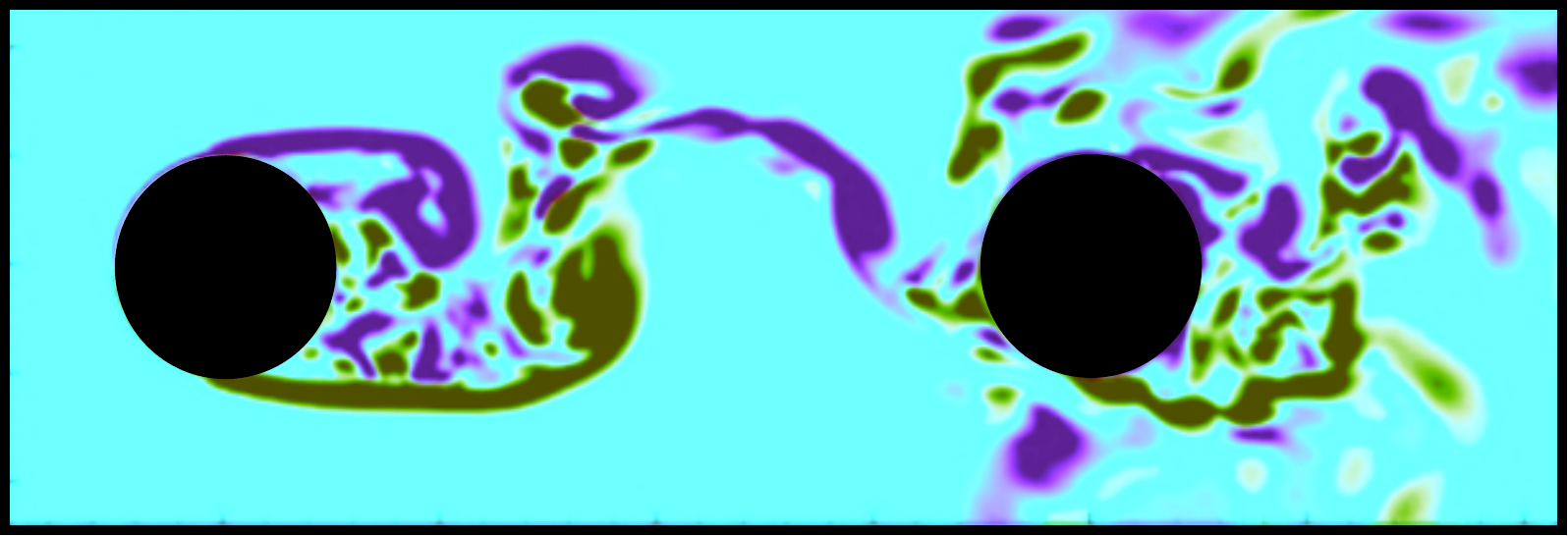}
\caption{$\ell/\mathcal{D}=4$}
\label{fig:midspanvortanc}
\end{subfigure}%
\caption{Normalized spanwise vorticity $\omega_z\mathcal{D}/\mathfrak{u}_{1,\infty}$ in the mid-span plane ($z=\pi/2$ for different gaps.}
\label{fig:midspanvortan}
\end{figure}

Flow structure is first analyzed to explore the flow characteristic around the two tandem circular cylinders in the subcritical regime characterized by $\Rey=2.2\times 10^4$.
The normalized spanwise vorticity $\omega_z\mathcal{D}/\mathfrak{u}_{1,\infty}$ in the mid-span plane $z=\pi/2$ is displayed in Fig.~\ref{fig:midspanvortan}.
In the case of smaller spacing ratio $\ell/\mathcal{D}=2$ (cf. Fig.~\ref{fig:midspanvortana}), no vortices are observed in the wake of the front cylinder.
The two bodies behave like a single lengthened bluff body with vortex formation behind the downstream cylinder.
The shear layers which separate from upstream
cylinder reattach to the surface of the downstream one, and then roll up into mature vortices behind the downstream cylinder, which results in a quasi-two-dimensional flow pattern characteristics.
For $\ell/\mathcal{D}=3$, the shear layer reattachment and vortex formation seem to coexist in the gap region as As portrayed in Fig.~\ref{fig:midspanvortanb}.
When $\ell/\mathcal{D}$ is increased to 4 as shown in Fig.~\ref{fig:midspanvortanc}, vortex shedding from both cylinders can be detected and one can notice two K\'arm\'an vortex streets occurring from the upstream cylinder as well as the downstream cylinder, and these vortices shed from the upstream cylinder impinge on the downstream cylinder.
For the small gaps ($\ell/\mathcal{D}$, the repulsive force between the cylinders is intense, which makes the wake clearly not organized. 
As the gap between the two cylinders gets higher, the formation of two independent wakes can be observed as in isolated cylinders.
This behavior was also observed by \citet{meneghini2001numerical}.

\begin{figure}[ht!]
\centering
\begin{subfigure}[h]{0.99\textwidth}
\includegraphics[width=0.99\textwidth]{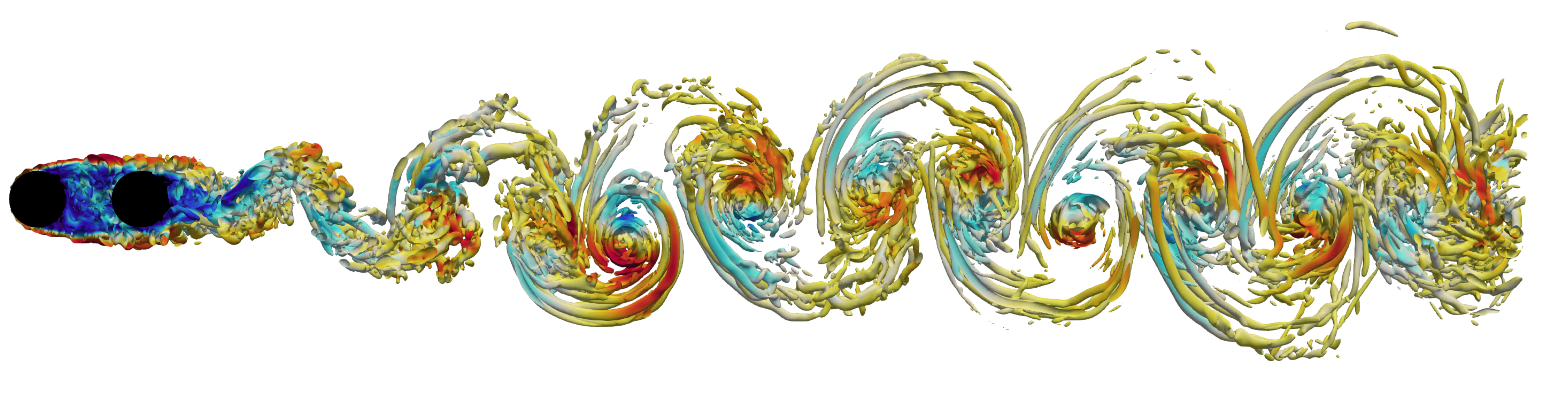}
\caption{$\ell/\mathcal{D}=2$}
\label{fig:vortana}
\end{subfigure}\\
\begin{subfigure}[h]{0.99\textwidth}
\includegraphics[width=0.99\textwidth]{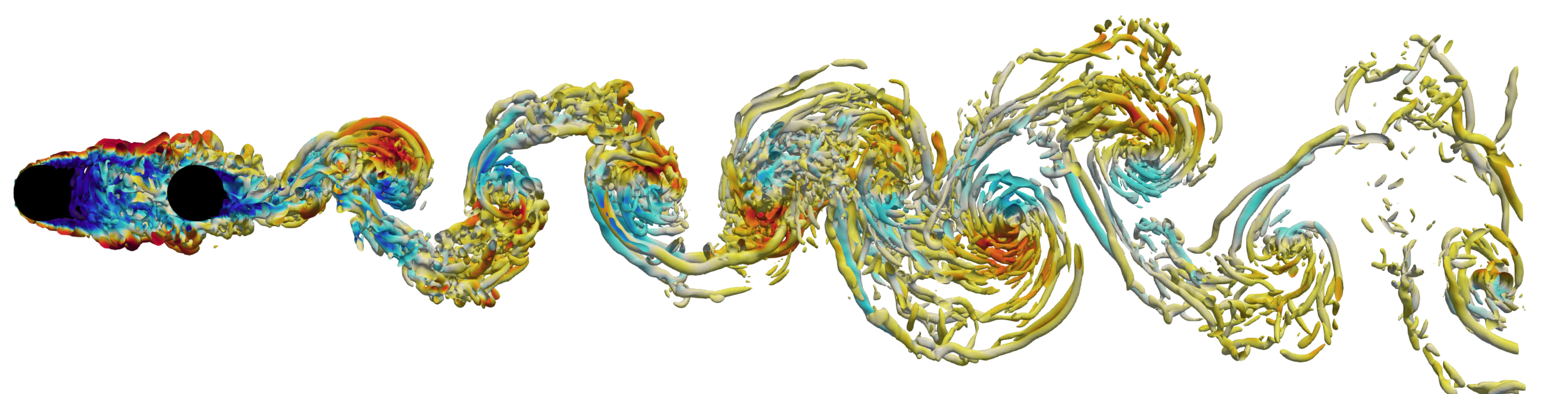}
\caption{$\ell/\mathcal{D}=3$}
\label{fig:vortanb}
\end{subfigure}\\
\begin{subfigure}[h]{0.99\textwidth}
\includegraphics[width=0.99\textwidth]{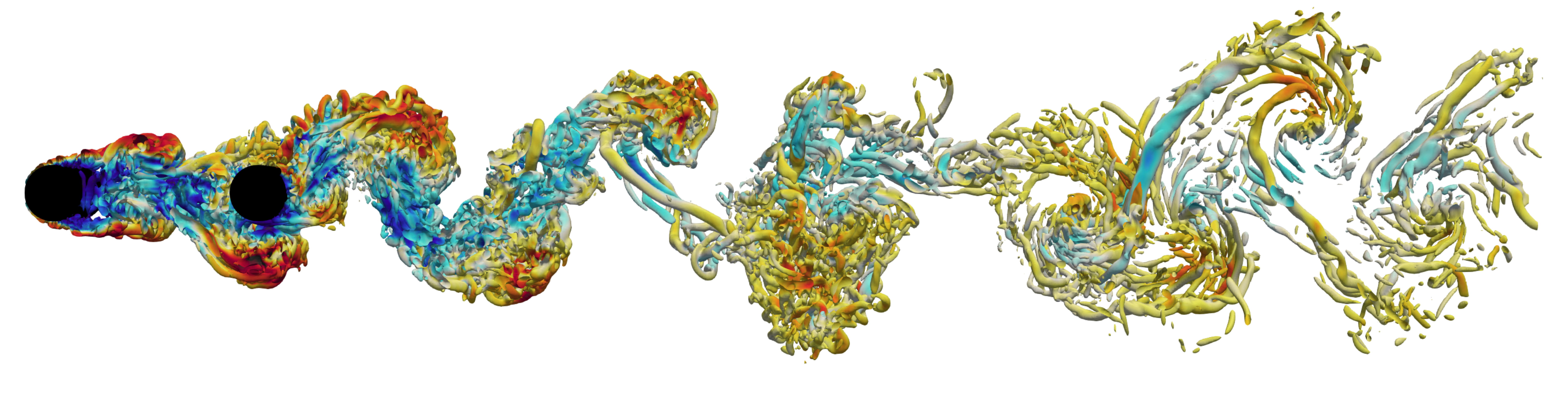}
\caption{$\ell/\mathcal{D}=4$}
\label{fig:vortanc}
\end{subfigure}%
\caption{Instantaneous snapshot of $Q$-criterion for tandem arrangement cylinders for different gaps.}
\label{fig:vortan}
\end{figure}

To get a deeper understanding of the three-dimensional vortical structures occurring in the present setup, instantaneous vortical structures represented by the isosurface of a $Q$-criterion are shown in Fig.~\ref{fig:vortan}, which presents the detailed instantaneous isosurface of $Q=1.5\mathfrak{u}^2_{1,\infty}/\mathcal{D}^2$.
As the gap between the two cylinders increases, the rib-like vortices dominate the wake flow, leading to
the increase of enstrophy of the flow fields and a greatly distorted wake.
Similar observations were made by \citet{dehkordi2011numerical} and \citet{tong2015numerical}.

\begin{figure}[ht!]
\centering
\begin{subfigure}[h]{0.32\textwidth}
\includegraphics[width=0.99\textwidth]{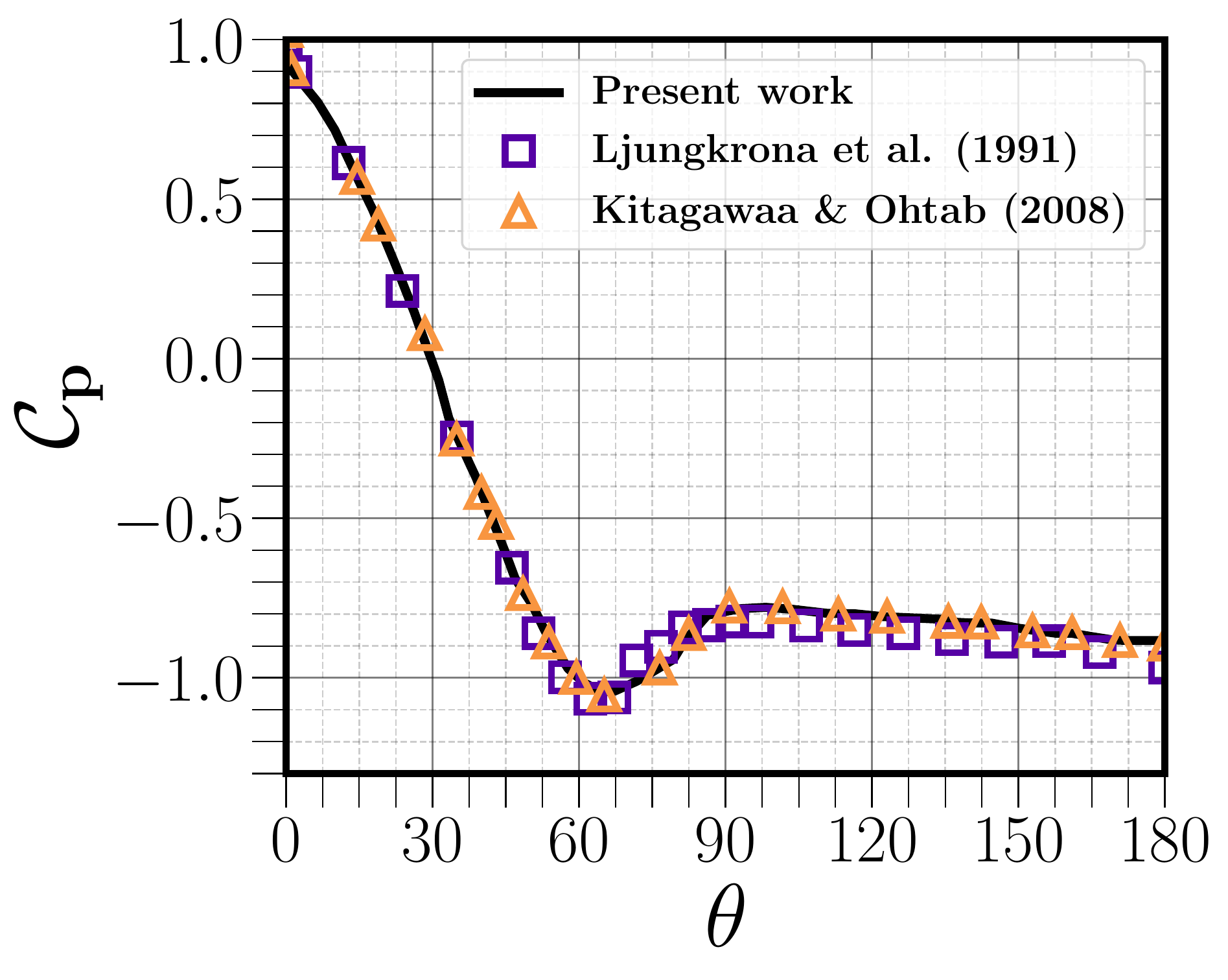}
\caption{$\ell/\mathcal{D}=2$}
\label{fig:cpdisa}
\end{subfigure}
\begin{subfigure}[h]{0.32\textwidth}
\includegraphics[width=0.99\textwidth]{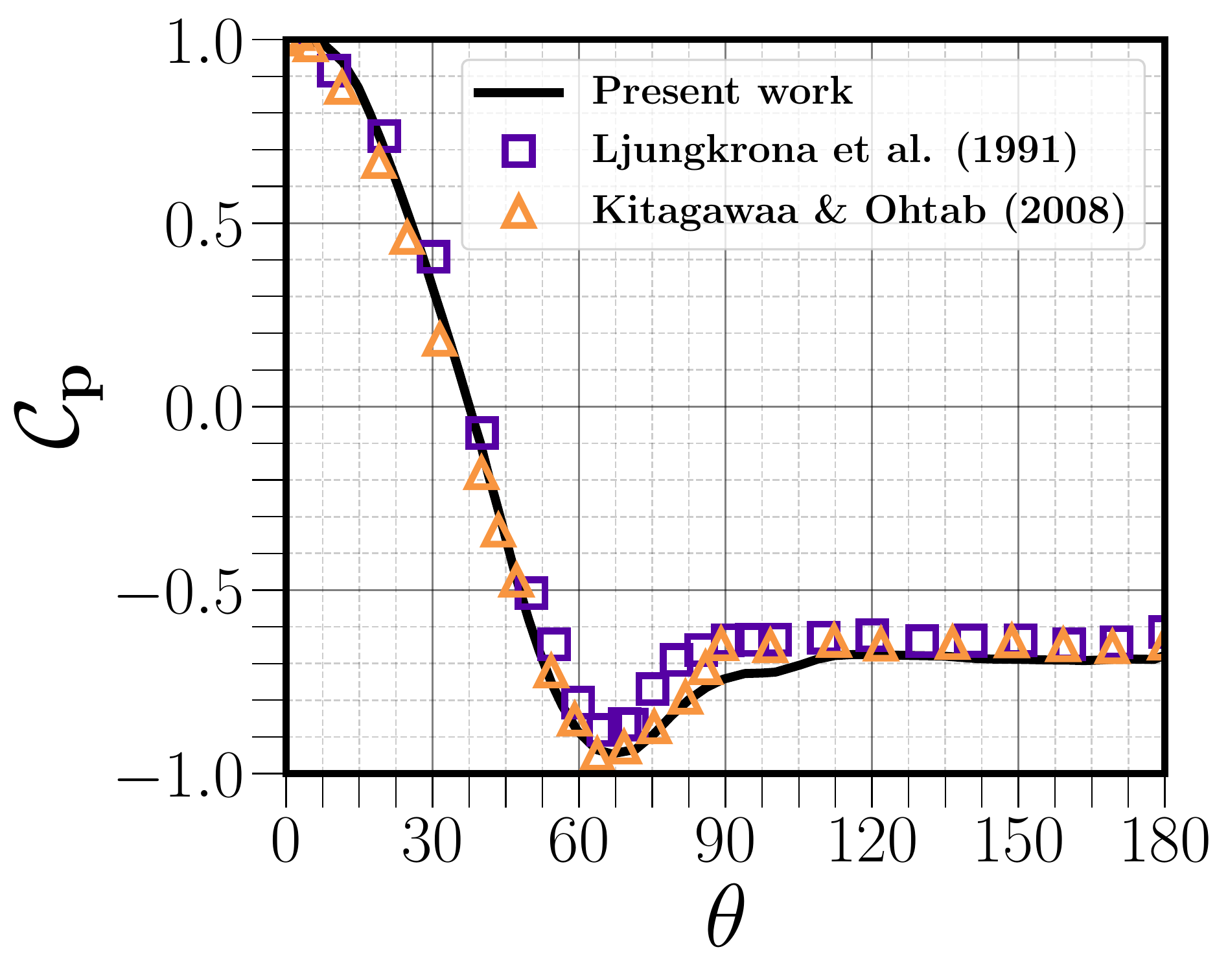}
\caption{$\ell/\mathcal{D}=3$}
\label{fig:cpdisc}
\end{subfigure}%
\begin{subfigure}[h]{0.32\textwidth}
\includegraphics[width=0.99\textwidth]{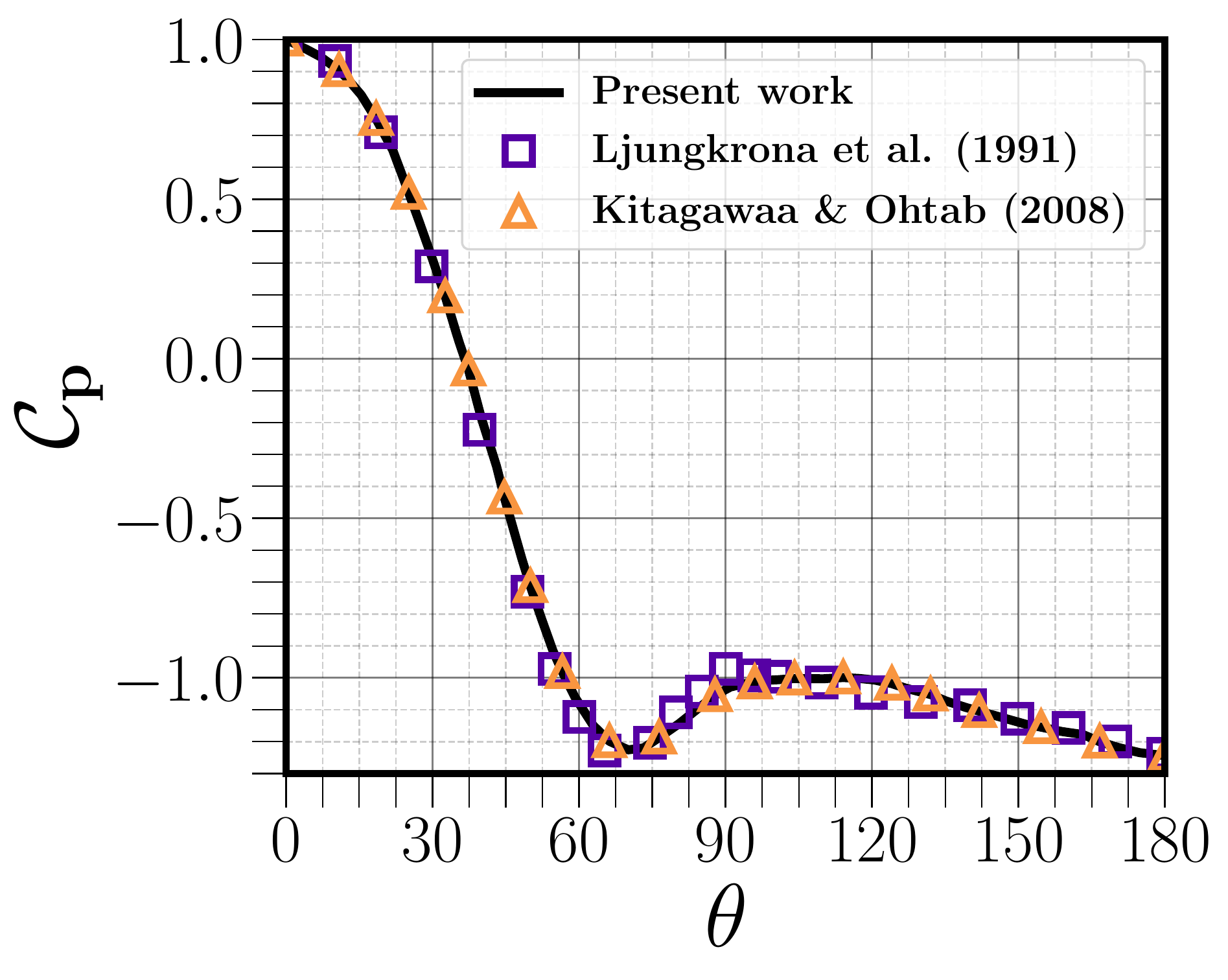}
\caption{$\ell/\mathcal{D}=4$}
\label{fig:cpdise}
\end{subfigure}\\
\begin{subfigure}[h]{0.32\textwidth}
\includegraphics[width=0.99\textwidth]{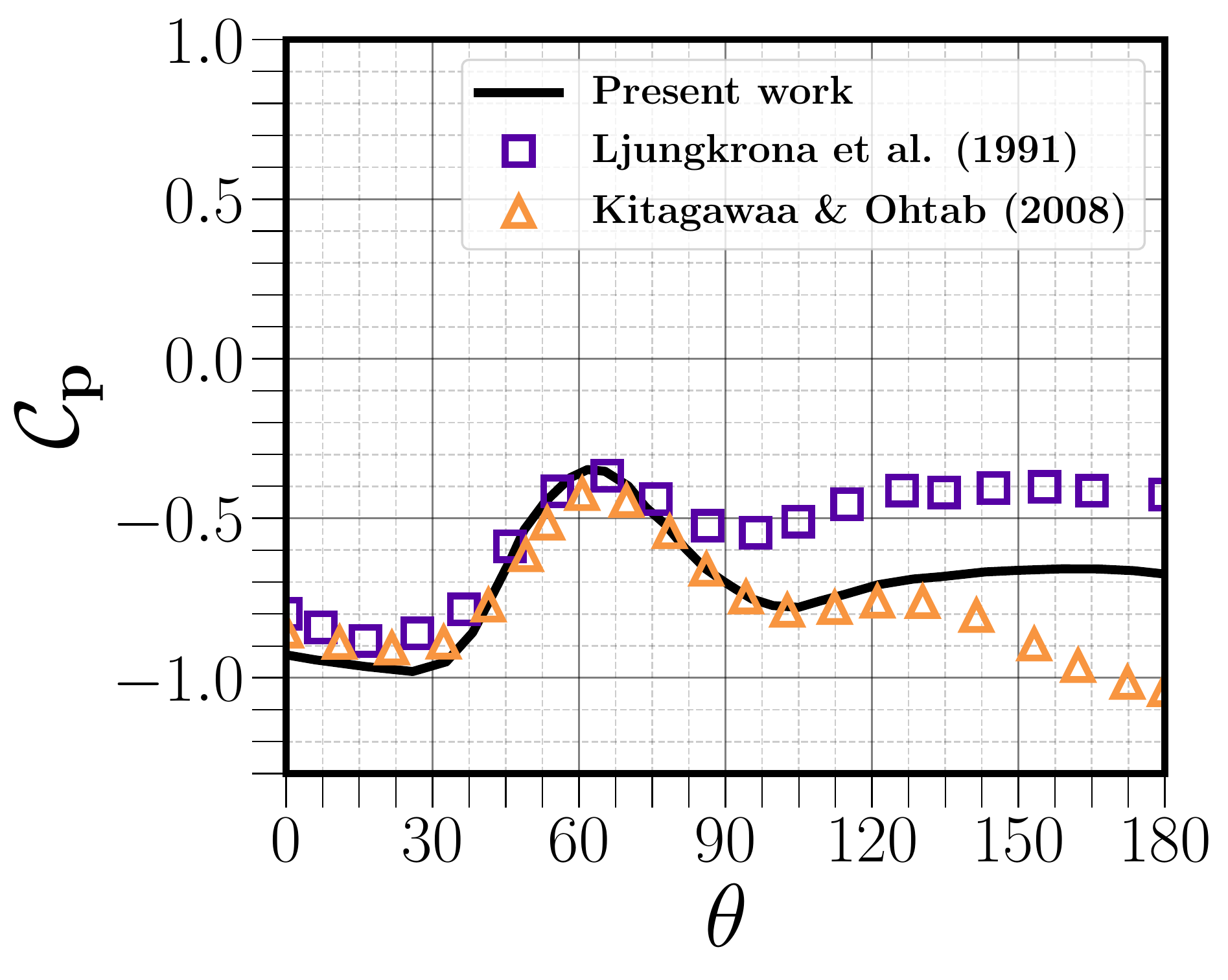}
\caption{$\ell/\mathcal{D}=2$}
\label{fig:cpdisb}
\end{subfigure}
\begin{subfigure}[h]{0.32\textwidth}
\includegraphics[width=0.99\textwidth]{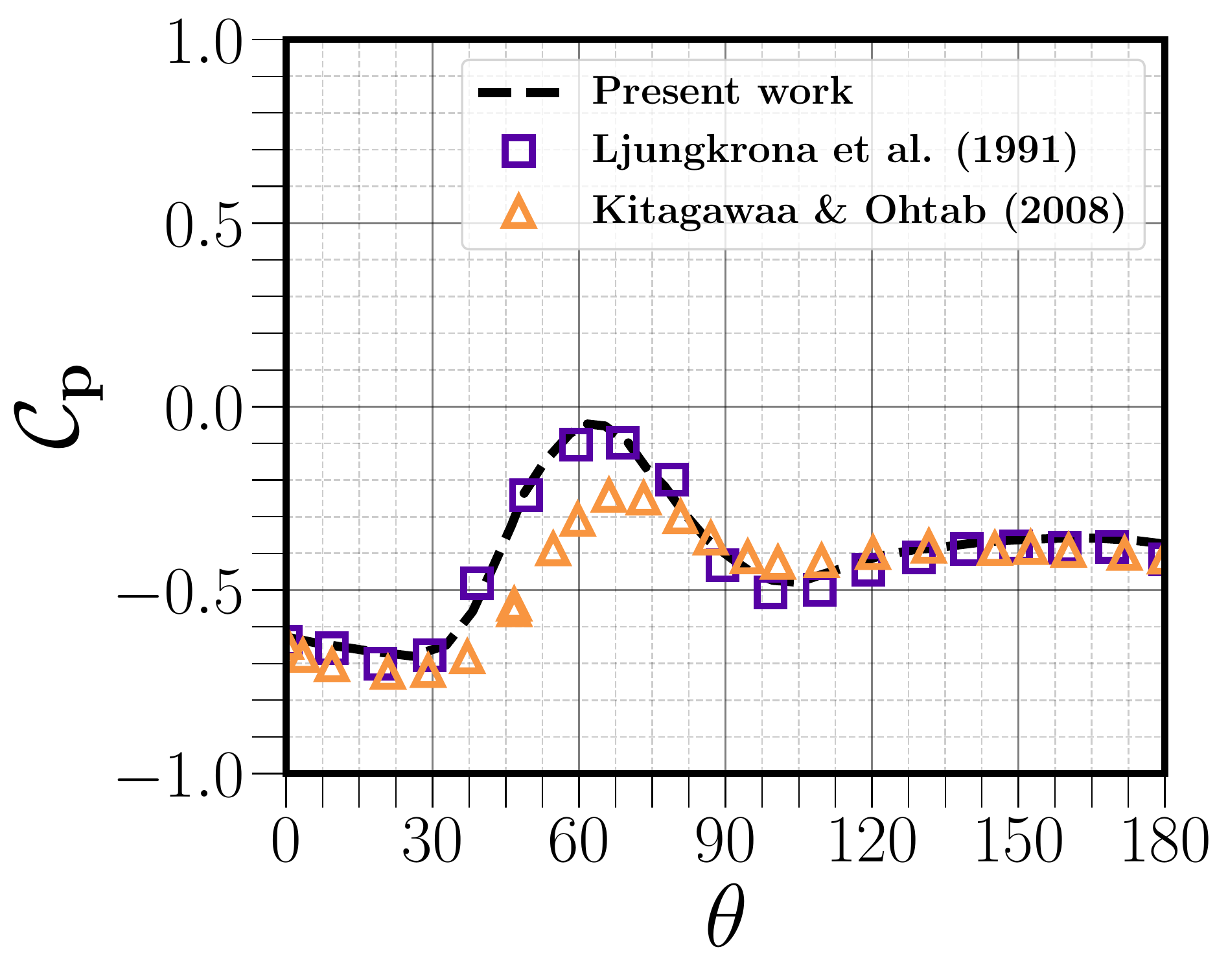}
\caption{$\ell/\mathcal{D}=3$}
\label{fig:cpdisd}
\end{subfigure}
\begin{subfigure}[h]{0.32\textwidth}
\includegraphics[width=0.99\textwidth]{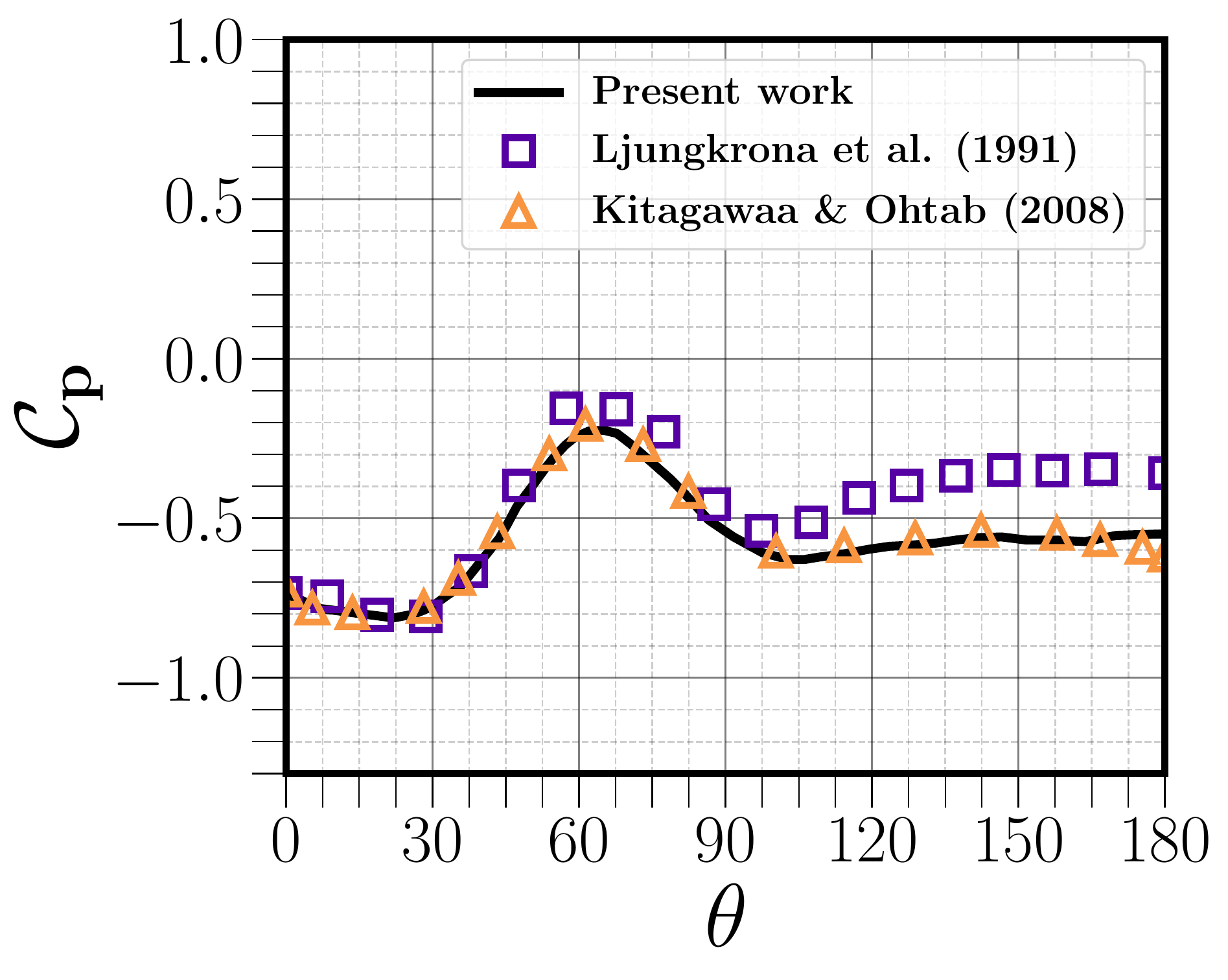}
\caption{$\ell/\mathcal{D}=4$}
\label{fig:cpdisf}
\end{subfigure}%
\caption{Time-averaged pressure distributions on surfaces of the upstream cylinder (top row) and downstream cylinder (bottom row) with different gaps in comparison with the experimental results at $\Rey= 2\times 10^4$ of \citet{ljungkrona1991free} and the numerical results at $\Rey= 2.2\times 10^4$ of \citet{kitagawa2008numerical}. 
}
\label{fig:cpdis}
\end{figure}

The ability of the present method to properly capture the surface pressure coefficients $\mathcal{C}_p=2(\mathfrak{p}-\mathfrak{p}_{\infty})/\varrho_{\infty}\mathfrak{u}_{1,\infty}^2$, where $\mathfrak{p}_{\infty}$ and $\varrho_{\infty}$ denote asymptotic physical quantities imposed at the inlet, as a function of the distance between their centers is shown in Fig.~\ref{fig:cpdis}.
It is observed that the distributions of $\mathcal{C}_p$ of upstream cylinder are almost unchanged when increasing the gap from $2$ to $3$.
However, for $\ell/\mathcal{D}=4$, the distribution of $\mathcal{C}_p$ rapidly decreases for $\theta\ge 100$, which is compatible with the observations of Fig.~\ref{fig:vortanc} since the larger vorticity magnitude is generated near the rear of the upstream cylinder.
For the downstream cylinder, the shape of $\mathcal{C}_p$ is quite similar for $\ell/\mathcal{D}=2$ an $3$. 
However, the case $\ell/\mathcal{D}=4$ exhibits a distribution of $\mathcal{C}_p$ that is always negative, which is due to the fact that the vortices shed from the upstream cylinder strongly impinge on the surface of the downstream cylinder.
The good agreement that is observed between the present computation and the results available in the literature,  verifies the correctness of the level-set immersed boundary method.

\begin{figure}[ht!]
\centering
\begin{subfigure}[h]{0.32\textwidth}
\centering
\includegraphics[width=0.99\textwidth]{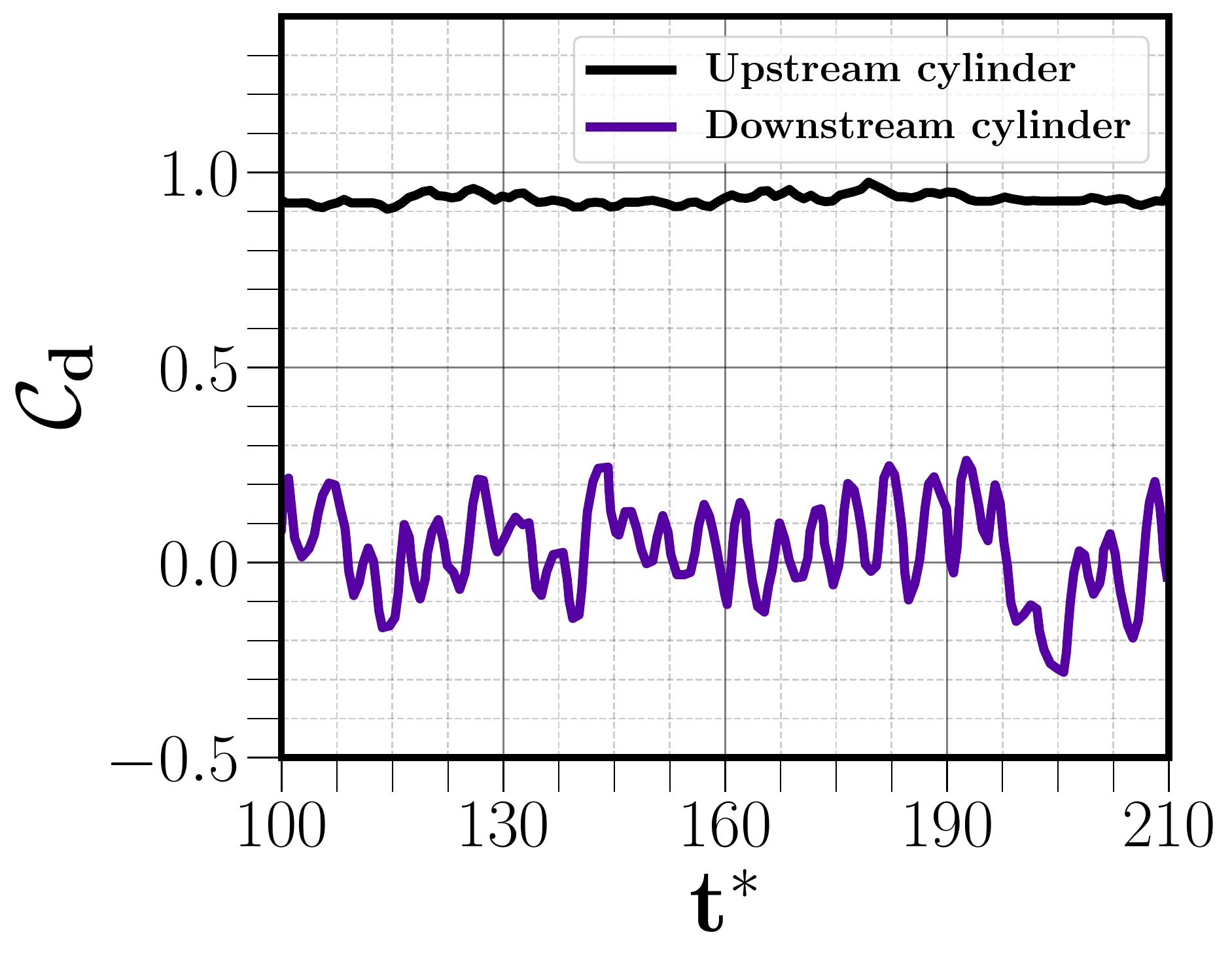}
\caption{$\ell/\mathcal{D}=2$}
\label{fig:cdhista}
\end{subfigure}
\begin{subfigure}[h]{0.32\textwidth}
\centering
\includegraphics[width=0.99\textwidth]{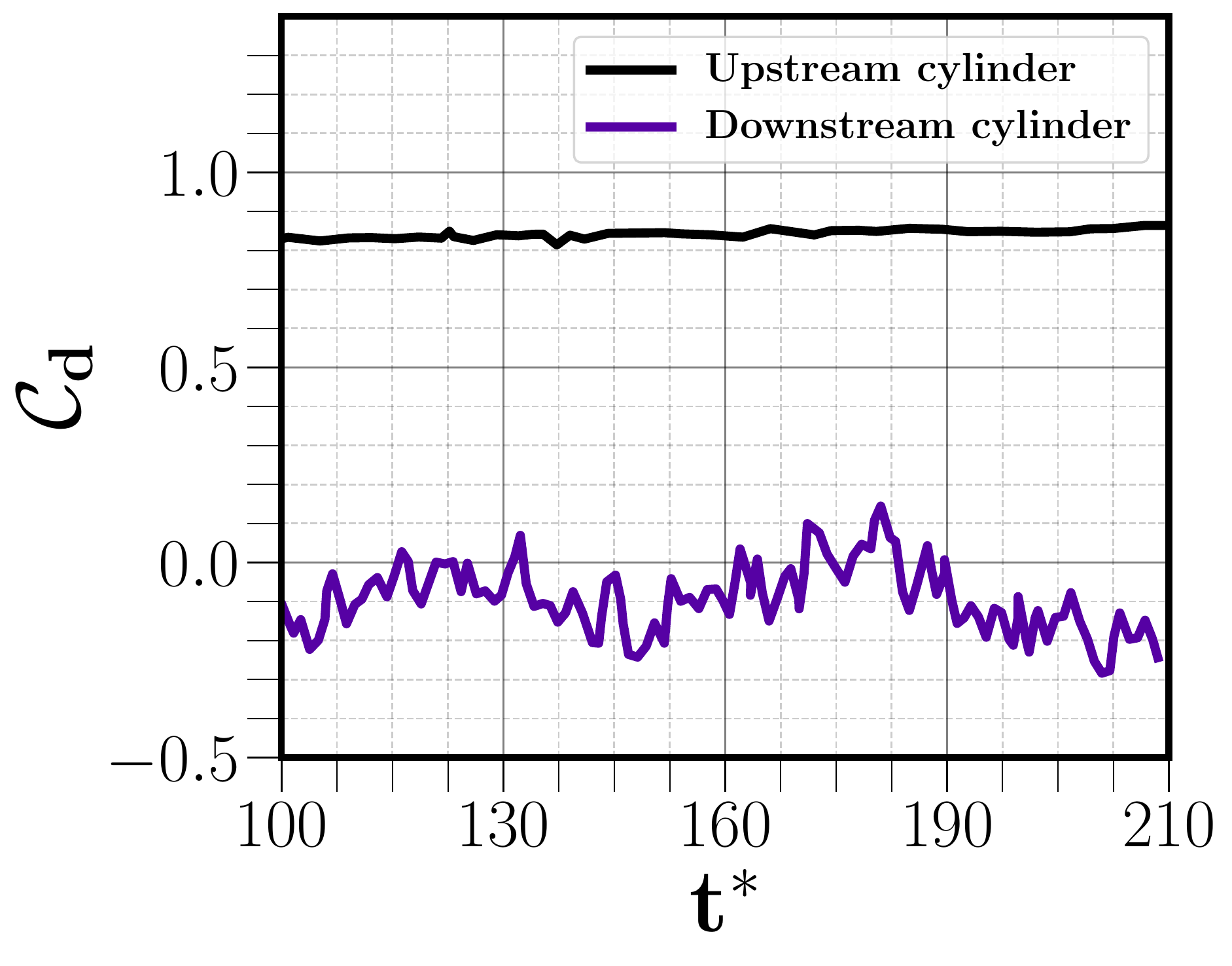}
\caption{$\ell/\mathcal{D}=3$}
\label{fig:cdhistb}
\end{subfigure}%
\begin{subfigure}[h]{0.32\textwidth}
\centering
\includegraphics[width=0.99\textwidth]{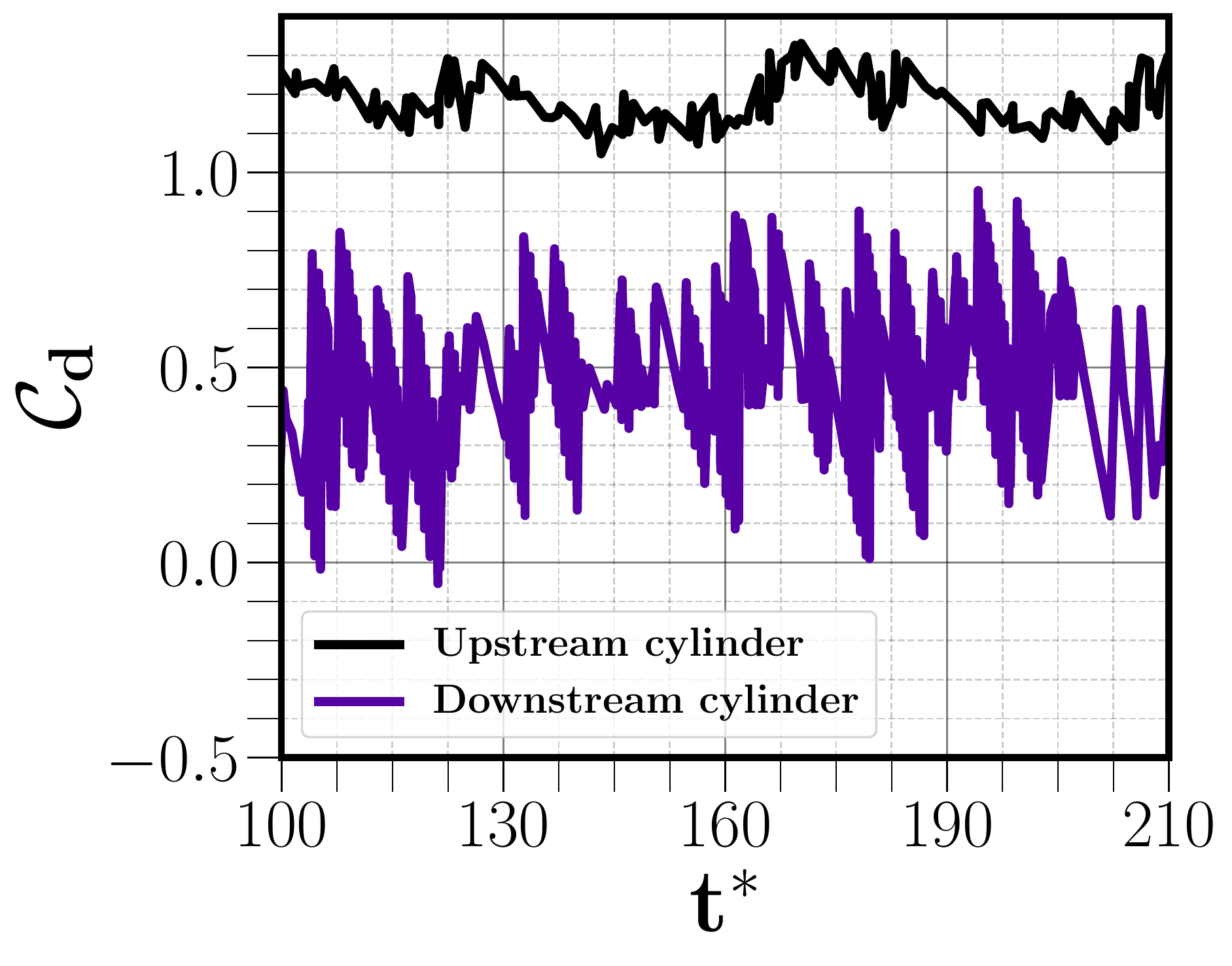}
\caption{$\ell/\mathcal{D}=4$}
\label{fig:cdhistc}
\end{subfigure}\\
\begin{subfigure}[h]{0.32\textwidth}
\centering
\includegraphics[width=0.99\textwidth]{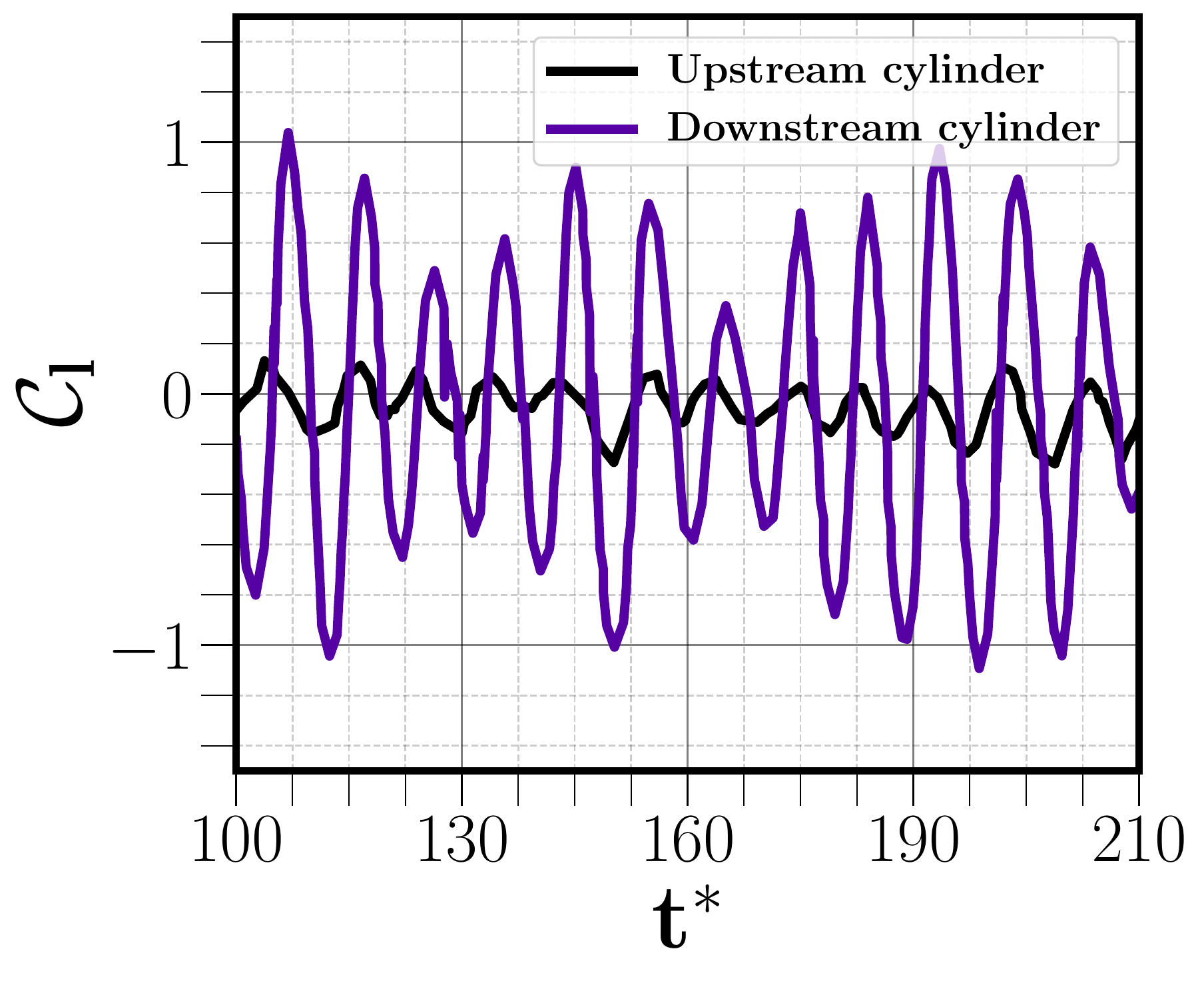}
\caption{$\ell/\mathcal{D}=2$}
\label{fig:cdhista}
\end{subfigure}
\begin{subfigure}[h]{0.32\textwidth}
\centering
\includegraphics[width=0.99\textwidth]{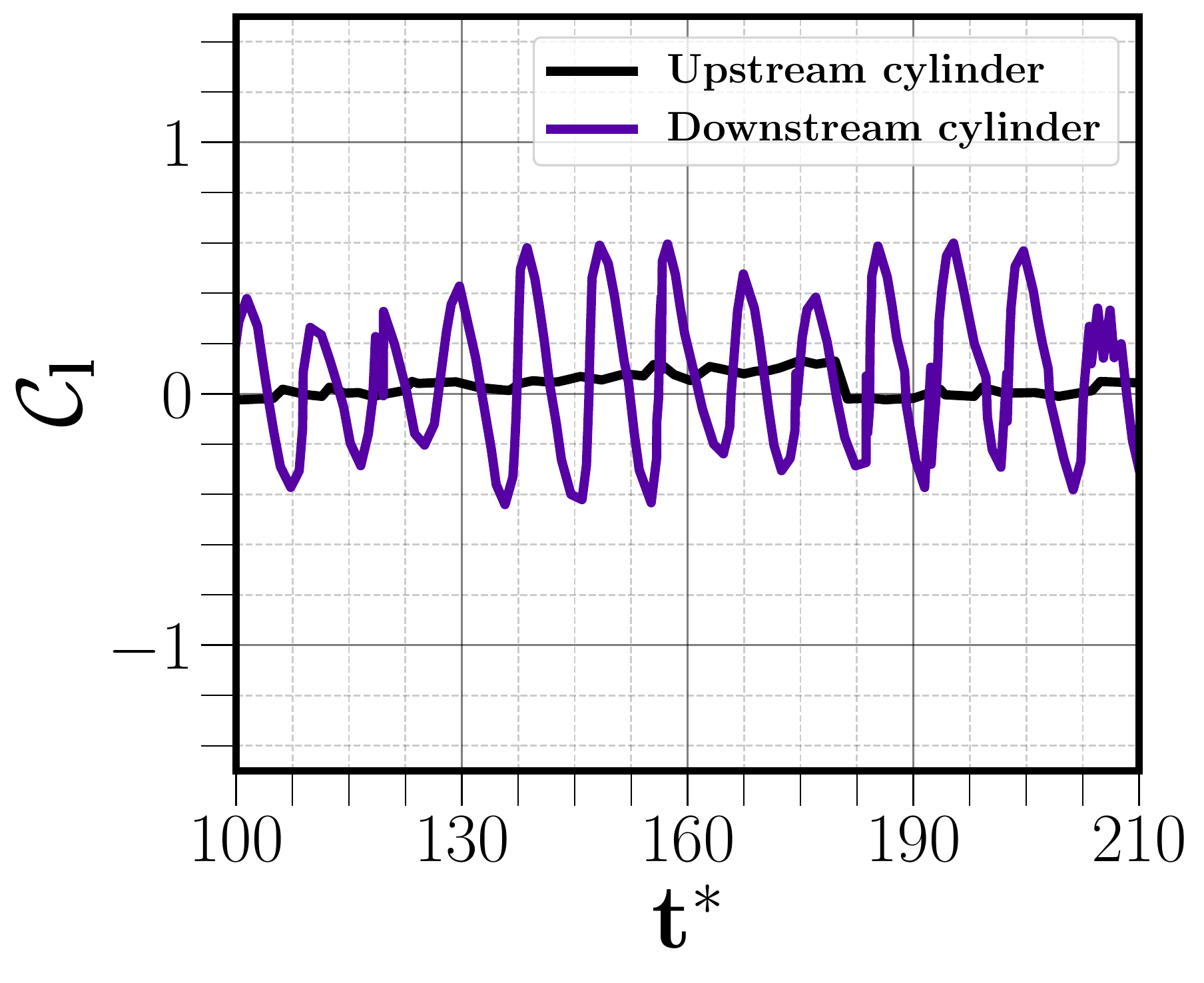}
\caption{$\ell/\mathcal{D}=3$}
\label{fig:cdhistb}
\end{subfigure}%
\begin{subfigure}[h]{0.32\textwidth}
\centering
\includegraphics[width=0.99\textwidth]{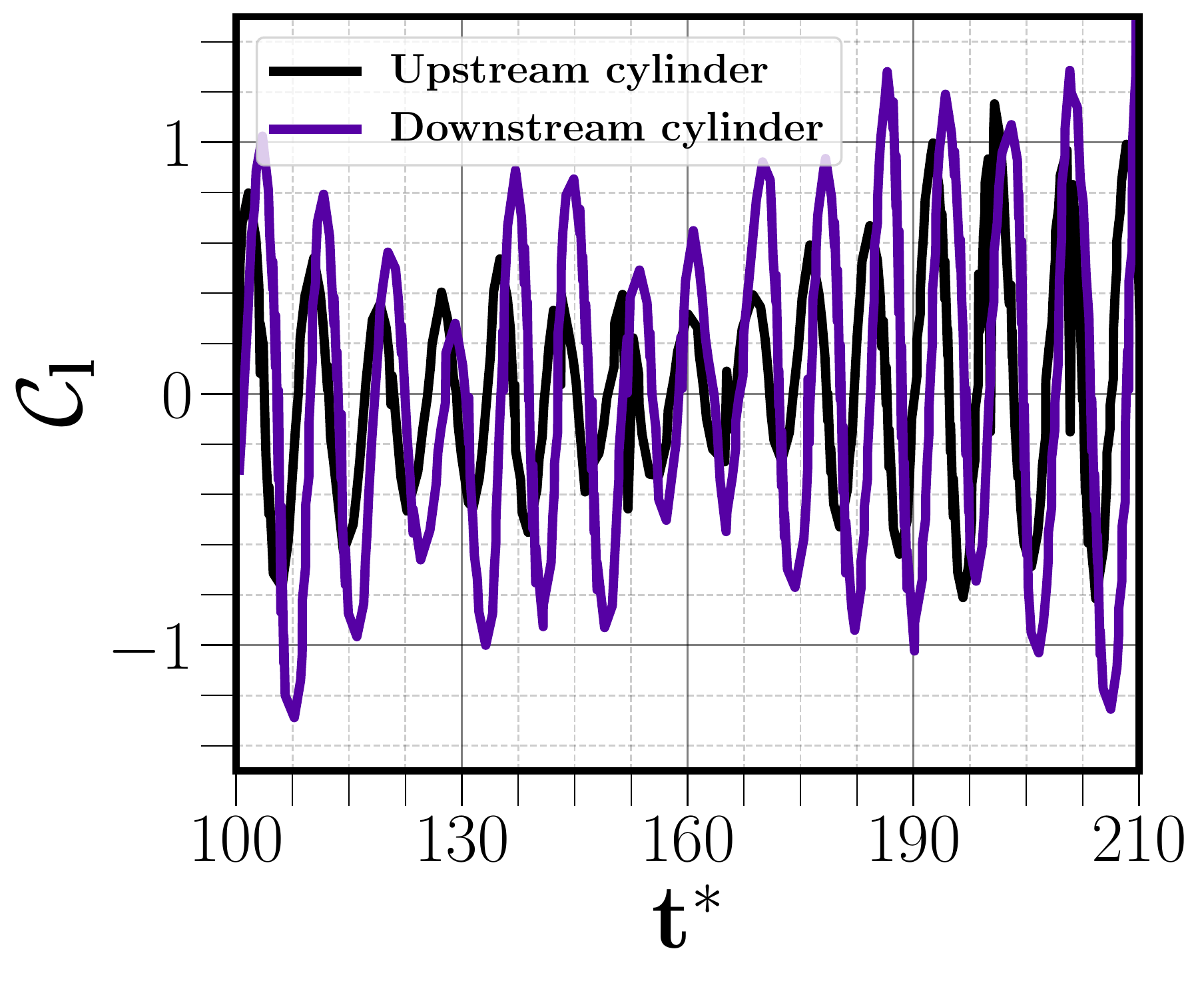}
\caption{$\ell/\mathcal{D}=4$}
\label{fig:cdhistc}
\end{subfigure}
\caption{Time history of drag and lift coefficients of the cylinders for different gaps.}
\label{fig:cdhist}
\end{figure}

The time histories of the drag coefficient $\mathcal{C}_d$ for different gaps is depicted in Fig.~\ref{fig:cdhist}.
Since no vortex is forming from the upstream cylinder, the drag coefficient fluctuations of the upstream are weaker than of the downstream cylinder for all gaps.
These fluctuations get stronger as the gap between the two cylinders becomes higher because the vortices shed from the upstream cylinders had various level of strength and are impinging the downstream cylinder surface at different positions.
Similar fluctuations are observed for $\mathcal{C}_l$, which exhibits quite different fluctuation magnitudes per vortex shedding period.
These observations were found in the study on the
tandem cylinders by \citet{kitagawa2008numerical} and \citet{gopalan2015numerical}.

The average calculation of the drag coefficients and the Strouhal number $\mathrm{S}t$, which is obtained from the power spectrum of the downstream-cylinder left coefficient, along with the experimental result at $\Rey= 2\times 10^4$ of \citet{ljungkrona1991free} and the numerical results at $\Rey= 2.2\times 10^4$ of \citet{kitagawa2008numerical} are reported in Tables.~\ref{tab:cdcomp} and \ref{tab:stcom}.
The most striking finding is that the drag coefficient sign of the downstream cylinder switches from a negative value to a positive value when the vortices of the upstream cylinder start to collide with the downstream cylinder, typically when $\ell/\mathcal{D}=4$ in our study.
One can notice also the strong dependence of the Strouhal number on the spacing.
For the average drag coefficient and Strouhal number, the
present results are in excellent agreement with data from the literature.
Finally, this last set of results confirms the capability of the present level-set immersed boundary approach to deal
with flows over bluff bodies and this offers quite encouraging perspectives for future applications of the proposed immersed boundary methodology.
\begin{table}[ht!]
\begin{tabular}{l|c|c|c||c|c|c}
\cline{2-7}
& \multicolumn{3}{c||}{Upstream cylinder} & \multicolumn{3}{c}{Downstream cylinder} \\ \hline\hline
$\ell/\mathcal{D}$ & 2 & 3  & 4 & 2 & 3 & 4 \\ \hline\hline
Present results & 0.882 & 0.807 & 1.180   &  -0.010 & -0.225 & 0.462\\
\citet{ljungkrona1991free} & 0.951 & 0.857 & 1.217 & -0.397 & -0.073 & 0.431\\
\citet{kitagawa2008numerical} & 0.895 & 0.819 & 1.200 & -0.009 & -0.212 & 0.453\\ \hline
\end{tabular}
\caption{Comparison of drag coefficient between the present computations, the experimental result at $\Rey= 2\times 10^4$ of \citet{ljungkrona1991free} and the numerical results at $\Rey= 2.2\times 10^4$ of \citet{kitagawa2008numerical}.}
\label{tab:cdcomp}
\end{table}
\begin{table}[ht!]
\centering
\begin{tabular}{l|c|c|c}
\cline{1-4}
$\ell/\mathcal{D}$ & 2 & 3  & 4 \\ \hline\hline
Present results & 0.160 & 0.156 & 0.184\\
\citet{ljungkrona1991free} & 0.167 & 0.144 & 0.177
\\
\citet{kitagawa2008numerical} & 0.161 & 0.154 & 0.186\\ \hline
\end{tabular}
\caption{Comparison of the Strouhal number between the present computations, the experimental result at $\Rey= 2\times 10^4$ of \citet{ljungkrona1991free} and the numerical results at $\Rey= 2.2\times 10^4$ of \citet{kitagawa2008numerical}.}
\label{tab:stcom}
\end{table}
\section{Conclusion}
%
In this paper, the accurate conservative level set with an important improvement of the modification of the re-initialization step is used for the two-way coupling of a fluid with rigid bodies.
Two cylinder in tandem arrangement has been done to investigate the performance of the method. 
The results of the mean drag and lift coefficients and the Strouhal number were compared with other author’s results and our calculations showed good agreement and offer quite promising perspectives for future applications to more complex industrial problems.
\section*{Acknowledgements}
The authors gratefully acknowledge support and computing resources from the African Supercomputing Center (ASCC) at UM6P (Morocco).
\bibliography{article.bib}
\end{document}